\documentclass[conference]{IEEEtran}
\usepackage{graphicx} 
\usepackage{textgreek}
\usepackage{amsmath,amssymb,amsfonts}
\usepackage{algorithmic}
\usepackage{textcomp}
\usepackage{xcolor}
\usepackage{soul}
\usepackage{listings}
\usepackage{url}
\usepackage[hidelinks]{hyperref}
\usepackage[utf8]{inputenc}
\usepackage[small]{caption}
\usepackage{booktabs}
\usepackage{adjustbox}
\usepackage{tabularx}
\usepackage{diagbox}
\usepackage[backend=biber, natbib=true, style=numeric, urldate=long, sorting=none, dateabbrev=false]{biblatex}
\title{Investigation of FlexAlgo for User-driven Path Control}

\makeatletter
\newcommand{\linebreakand}{
  \end{@IEEEauthorhalign}
  \hfill\mbox{}\par
  \mbox{}\hfill\begin{@IEEEauthorhalign}
}
\makeatother

\author{\IEEEauthorblockN{1\textsuperscript{st} Julia Kułacz}
\IEEEauthorblockA{\textit{Multiscale Networked Systems group} \\
\textit{University of Amsterdam}\\
Amsterdam, The Netherlands \\
julia.kulacz@os3.nl}
\and
\IEEEauthorblockN{2\textsuperscript{nd} Martyna Pawlus}
\IEEEauthorblockA{\textit{Multiscale Networked Systems group} \\
\textit{University of Amsterdam}\\
Amsterdam, The Netherlands \\
martyna.pawlus@os3.nl}
\linebreakand 
\IEEEauthorblockN{3\textsuperscript{rd} Leonardo Boldrini}
\IEEEauthorblockA{\textit{Multiscale Networked Systems group} \\
\textit{University of Amsterdam}\\
Amsterdam, The Netherlands \\
l.boldrini@uva.nl}
\and
\IEEEauthorblockN{4\textsuperscript{th} Paola Grosso}
\IEEEauthorblockA{\textit{Multiscale Networked Systems group} \\
\textit{University of Amsterdam}\\
Amsterdam, The Netherlands \\
p.grosso@uva.nl}
}

\date{March 2023}

\begin{document}

\maketitle

\begin{abstract}
This paper examines the Flexible Algorithm (FlexAlgo) for its potential to enable user-driven path control in intra-domain Segment Routing (SR) enabled networks. FlexAlgo is a relatively new approach to intra-domain routing that allows multiple custom algorithms to coexist within a single domain. This capability has the potential to provide users with greater control over the paths their data takes through a network. The research includes a thorough investigation of the FlexAlgo approach, including an examination of its underlying techniques, as well as a practical implementation of a FlexAlgo-based solution. We depict performed experiments where we implemented FlexAlgo in three different scenarios. We also present how we developed an automated tool for users to control traffic steering using preferred metrics and constraints. The results of this investigation demonstrate the capabilities of FlexAlgo as a means of enabling user-driven path control and therefore increase security and trust of users towards the network.
\end{abstract}

\section{Introduction}\label{sec:intro}
The need to ensure trustworthiness and privacy to network users has led to developments such as the Responsible Internet \cite{hesselman2020responsible}. One of the still open questions has been which technologies can be used to support user-driven path control.  

For many years, the Interior Gateway Routing Protocol (IGP) was the primary factor in determining the optimal paths across intra-domain networks. This solution was rather limited as it takes solely bandwidth into account when calculating the best paths throughout the network. However, as network architectures have become more complex and dynamic, routing has become much more sophisticated thanks to Traffic Engineering (TE) solutions. It started taking into consideration a wider range of factors beyond just bandwidth. Therefore, the TE solutions allow network operators to manage the traffic in a way that meets the specific needs and requirements of the network and also its users. One aspect of TE is user-driven path control, which focuses on putting control over some decisions, related to choosing the best paths, in the hands of the users. This allows them to specify different metrics or constraints that the network should use to determine the optimal path for their traffic. By allowing this level of control over the network to its users we increase their situational awareness over how  their traffic is being routed. This brings an additional layer of security, as users can choose paths that are the most secure specifically for their traffic. It also increases their trust towards the network as they are aware and monitor what happens to their data.

In this paper, we aim to implement and test one of the newly introduced TE solutions called Flexible Algorithm (FlexAlgo). Moreover, we want to examine which metrics and constraints can be used in order to give users control over their traffic. This falls into the scope of UPIN (User-driven Path verification and control for Inter-domain Networks) as the goal of this project is to give users more insight and control over their data in transit through the network, to finally increase the security and trustworthiness of users to the network. These considerations led us to state the main research question of this paper which is: what are the capabilities and limitations of FlexAlgo when used to support user-driven path control?



\section{Background} \label{Background}


\subsection{Segment Routing}
The underlying technique that we use in order to provide user-driven path control is Segment Routing (SR). SR is based on the source routing paradigm where packets are steered through a set of instructions called the SR Policy. The SR Policy consists of segments identified by their Segment IDs (SIDs) which can be associated with various types of topological or service instructions. Segments have a meaning that is either local to an SR node or global within the SR domain. The set of local segments is called SR Local Block (SRLB) and the set of global segments within an SR domain is called SR Global Block (SRGB).

Furthermore, SR can be implemented on different types of data planes, the two most common ones being SR over MPLS (SR-MPLS) and SR over IPv6 (SRv6). In SR-MPLS, a segment is represented as an MPLS label and SR is implemented without any changes to the forwarding plane. The experiments conducted during our research concentrate on the SR-MPLS data plane and the Interior Gateway Protocol (IGP) based control plane. There are two different segments specified in the IGP-based distributed control plane, namely the IGP-Adjacency segment and the IGP-Prefix segment. The IGP-prefix segment represents a global instruction to forward traffic based on the specific IGP routing protocol and its SID is referred to as Prefix-SID. The IGP-Adjacency Segment is usually a local segment that represents an instruction to forward a packet to the next-hop router in the SR domain and its SID is referred to as Adj-SID \cite{Filsfils_Previdi_Ginsberg_Decraene_Litkowski_Shakir_2018}.

One of the primary applications of SR is Segment Routing Traffic Engineering (SR-TE). The SR-TE architecture allows directing the traffic by specifying an SR policy. This policy uses segment lists to represent a desired path and steer the traffic through it.
An SR Policy is represented by three factors; headend, color, and endpoint. The headend is a node that instantiates an SR Policy (e.g. edge router), whereas the endpoint is the target of an SR policy, both identified as IP addresses. Color is employed to link a policy with a desired objective, such as minimizing delay.
The set of segment lists used in SR-TE can be acquired from intent-based SR-TE algorithms. One of the recently proposed SR-TE algorithms is Flexible Algorithm \cite{Wu_Cui_2022}.
A headend node of SR is responsible for steering the traffic into a particular SR policy. When a prefix that is colored with a BGP Extended Community matches an SR Policy, the packet with the corresponding prefix can be steered with On-Demand Next Hop (ODN) policy. The purpose of ODN is to dynamically instantiate an SR Policy to a BGP next-hop \cite{SR-TE_2017}. BGP Extended Community is a path attribute that allows to group destinations which share a common property.

\subsection{Flexible Algorithm}\label{fad}

The SR architecture utilizes FlexAlgo to achieve its goals of efficient and flexible forwarding of packets as FlexAlgo can be associated with a Prefix segment \cite{Wu_Cui_2022}.
In general, FlexAlgo is used to calculate the best path along a constraint topology, and as described in the draft RFC \cite{Psenak_Hegde_Filsfils_Talaulikar_Gulko_2022}, it can be applied as the extension to IGPs such as IS-IS, OSPFv2, and OSPFv3. Within this research paper, we focus on describing and implementing FlexAlgo for OSPFv2.

In terms of network configuration, FlexAlgo is a numeric identifier in the range between 128 and 255. Within this paper, we refer to this value as FlexAlgo ID. It specifies the number related to the Flexible Algorithm Definition (FAD) which is the most fundamental term related to FlexAlgo. FAD is defined as a set of 3 components, namely calculation type, metric type, and constraints.

\textit{Calculation type} can be equal to 0 or 1. When equal to 0, it utilizes the Dijkstra Shortest Path First (SPF) algorithm in a traditional IGP and allows local policies to change the paths calculated by the SPF. When it is equal to 1, a strict SPF is in use and no local policy is allowed to modify paths \cite{sr-routing}. In all our experiments described in Section \ref{experiments},
we use the calculation type of value 0.

\textit{Metric type} defines a metric that is used during the calculation of FlexAlgo. There are three values defined, namely IGP Metric, Traffic Engineering Default Metric, and Minimum Unidirectional Link Delay \cite{flex-algo}.

\textit{Constraints} do not have a strict definition, they can indicate for example using only a particular network plane, including only a certain metric, or avoiding specific links in a network topology. The only constraints that we will focus on in our experiments relate to the links.
Two attributes needed to define them are affinity map and administrative group. The affinity map defines colors and associates them with unique bit positions. These bit positions relate to the bitmask sent in the OSPF Link State Update (LSU) and specify which bits are required to match the colors set on links. To make use of the affinity maps, the links can be colored by referring to a specific affinity map using the corresponding color name. Then, the constraint can be defined using the second component which is the administrative group. It makes use of one of the three arguments, meaning exclude-any, include-all, and include-any where each of them can contain a maximum of 10 colors related to the affinity maps \cite{SR-TE}.

In order to advertise the definition of FlexAlgo and therefore have loop-free computed paths, routers use OSPF FAD TLV (Type-Length-Value).
Before a node calculates the best paths for a specific FlexAlgo, it has to fulfill two conditions: it must be configured to participate in this specific FlexAlgo and it must select a consistent FAD that corresponds to this FlexAlgo. This FAD needs to include metric type and calculation type, but constraints are not mandatory. The first step towards calculating the best path for a FlexAlgo is defining links that should be definitely excluded from the computation.
Then, when all links are either excluded or included from the topology for a given FlexAlgo, the SPF algorithm can be run over this topology with a metric defined in the FAD. When the calculation is done, the Prefix-SID for a given FlexAlgo is installed in the MPLS forwarding table.



\section{Setup description} \label{Methodology}
\begin{figure}[t!]
    \centering
    \resizebox{\columnwidth}{!}{
    \includegraphics{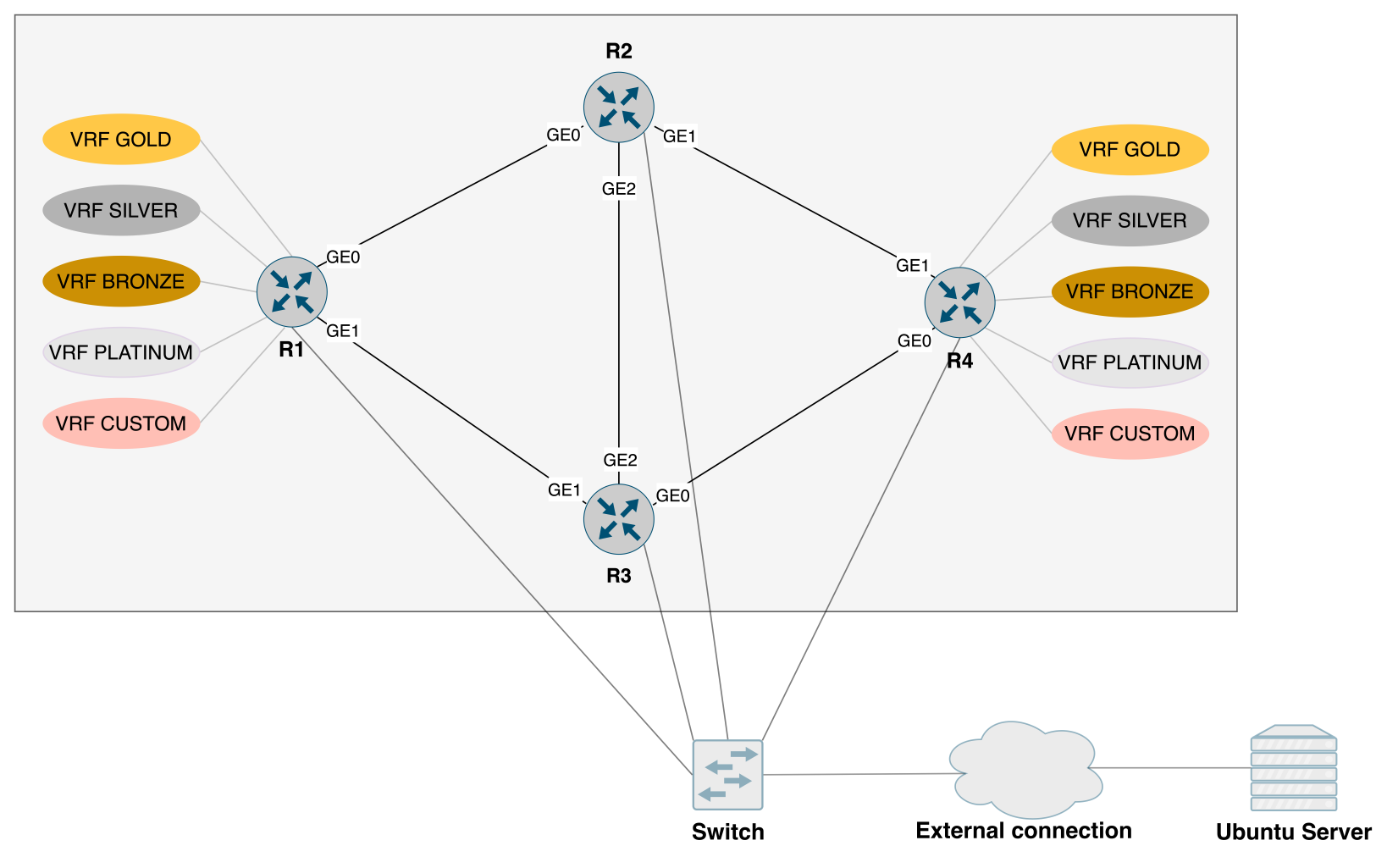}}
    \caption{Network topology created in Cisco Modelling Labs.}
    \label{fig:topology}
\end{figure}

We present the topology of our setup in Figure \ref{fig:topology}. It is divided into two logical parts. The first one is framed by the grey rectangle and poses the base for investigating FlexAlgo's capabilities. We used the additional three elements outside of this rectangle for automation purposes which are described in Section \ref{sub:pathcontroller}. As the base topology, we created a partial-mesh topology consisting of four XRv9000 nodes.

In order to connect the nodes, we used five GigabitEthernet links as presented in Figure \ref{fig:topology}.

Next step involved configuring the OSPF routing protocol on all nodes. We created the OSPF process on each router and assigned all interfaces to area 0 and defined them as OSPF point-to-point interfaces.

We configured SR in one domain where R1 executed ingress procedures, R4 executed egress procedures, whereas R2 and R3 were treated as transit nodes. We defined the SRGB in the global configuration so that each Prefix-SID was calculated from this base. We then activated SR-MPLS in all routers in area 0 of the OSPF routing protocol and configured all Loopback0 interfaces to calculate their Prefix-SIDs using the SRGB.

Lastly, in order to simulate traffic from five different users, we performed several steps on two edge routers, R1 and R4. Firstly, we configured BGP across SR-MPLS and created iBGP peering between R1 and R4. Next, we created five separate Virtual Routing and Forwarding (VRFs), namely GOLD, SILVER, BRONZE, PLATINUM, and CUSTOM, and we assigned five separate GigabitEthernet interfaces to these VRFs respectively. In general, VRFs are the layer 3 equivalent of layer 2 VLANs and allow to create multiple routing and forwarding tables within one router. The router itself stores more than one routing table, but the VRF instance allows using only one routing table, for a single user, at a time. To distinguish the VRFs from each other, we used Route Distinguisher (RD). RD is an identifier added to each router that identifies a specific VRF and distinguishes VRFs from each other. Then, we configured colors associated with five VRFs and advertised them in the BGP extended community attributes via Route Policy. This way, they were dynamically assigned to the VRFs later on. Afterwards, we configured the Route Policy in the outgoing direction, facing the other iBGP peer. In the last step, we created ODN policies for each color associated with a separate VRF. Within each policy, we specified the Prefix-SID of FlexAlgo associated with this specific VRF. All these steps allowed us to separate traffic coming from different users and are described in detail in Section \ref{experiments}.

As a part of the experiments, we also aimed to develop an automated method for a user to control traffic steering in the SR domain. This approach constitutes one possible implementation of the Path Controller in the UPIN project \cite{UPIN3}. In particular, our intention was to create an elementary user interface. It would allow users from VRF CUSTOM to either choose one of the existing FlexAlgos or create a new one that fulfilled their preferences. We implemented the controller using Ansible as a configuration management tool. The controller communicated with the routers using the NETCONF protocol \cite{netconf_rfc}. Moreover, to be able to retrieve and change configurations related to FlexAlgo and SR, we created specific RPC payloads with Cisco Yang Suite \cite{yang_rfc}. 



\section{Experiments}\label{experiments}


Before conducting experiments we had to configure fundamental techniques essential to test FlexAlgo capabilities. Firstly, we provided reachability between the routers by configuring the OSPF protocol. We added all routers into the backbone area, configured the interfaces as point-to-point networks, and enabled \textit{segment-routing mpls}. Then, we configured segment routing itself.

We then configured VRFs and BGP. The following configurations were only applied on router 1 and router 4 which were iBGP peers.
We created five VRFs, namely GOLD, SILVER, BRONZE, PLATINUM, and CUSTOM. Next, we assigned them route distinguishers, namely  1:1, 1:2, 1:3, 1:4, and 1:5 respectively. In order to  define which routes were exported and imported into a specific VRF routing table, we configured \textit{route-target import} and \textit{export}. We also assigned interfaces GigabitEthernet0/0/0/5-9 to the VRFs.


Finally, we configured the route policy and ODN. We created five BGP colors extended communities that we later applied to route distinguishers via route policy \textit{bronze\_silver}. At this point, we created five ODN SR policies with the colors mentioned earlier. We also associated the policies with different FlexAlgos.

\subsection{Experiment 1 - IGP metric and constraints}\label{sub:igp}
In Experiment 1, we created two FlexAlgo definitions 128 and 129 which were later associated with traffic from VRFs GOLD and SILVER respectively.

The affinity map associates the color red with a bit position 20 and blue with a bit position 10. These bit positions relate to the bitmask sent in the OSPF LSU in the Extended Admin Group field and specify which bits are required to match the colors set on links. Then, to make use of the affinity map, we colored the links by specifying colors on the routers' interfaces.
In the next step, we linked the configuration of FAD with the running OSPF process on all routers. To participate in the specific FlexAlgo and advertise its definition all routers had to configure it in the \textit{router ospf 1} submode.


Similarly, we configured FlexAlgo 129 on all routers.


The last remaining part of the configuration was the advertisement of Prefix-SID for a particular FlexAlgo. This way, all routers installed an MPLS-labeled path for the specific destinations.


From this point, we could see the advertisement of FlexAlgo 128, and 129 in the LSU packet in Link State Advertisement (LSA) type 10, particularly in Opaque Router Information LSA.
Within the exclude-any constraint, FlexAlgo 128 found matching to the color specified in the affinity map, meaning blue. Therefore, all links marked with the color blue are excluded from this FlexAlgo calculation. 


Once routers advertised the FAD, they computed the best paths to reach other nodes participating in specific FlexAlgos.




All routers computed their paths for the FlexAlgo 129 in a respective way. However, in this scenario, router 1 used the interface GigabitEthernet0/0/0/1 to reach the remaining nodes as red links were excluded from the computation. 

If we check the MPLS forwarding table on router 1 related to the Prefix-SIDs for FlexAlgo 128, we can see that, in order to forward traffic to the destination associated with Prefix-SID 20013 (which is a Prefix-SID for FlexAlgo 128 on router 3), router 1 would use the same outgoing label 20013. It would forward the traffic to interface GigabitEthernet0/0/0/0 to the Next Hop 10.0.12.2.


Finally, we associated FlexAlgo 128 and 129 with the previously configured ODN SR policies for colors 10 and 20 respectively.
This way, traffic generated within VRFs GOLD was forwarded via paths computed by FlexAlgo 128. 





\subsection{Experiment 2 - Delay metric}\label{sub:delay}

Experiment 2 focused on specifying the metric type as Minimum Unidirectional Link Delay for FlexAlgo 130 which we associated with VRF BRONZE. 
In order to measure delay on links and advertise it, we had to configure performance measurement on routers. In a real environment, routers would use performance measurement to monitor network metrics using probing mechanisms on the interfaces. However, in the CML virtualized environment  routers do not support performance measurement functionalities in the context of dynamic link delay measurement. In our scenarios, performance measurement allowed us to configure statically assigned delay values on the interfaces.


The routers advertised static delay values in the OSPFv2 Extended Link Opaque LSA of the OSPF LSU Packet. 
Router 1 advertises a Unidirectional Link Delay 100 \textmu s on the point-to-point link between router 1 and router 3. OSPFv2 Extended Link TLV defined the delay in Application-Specific Link Attributes Sub-TLV called Unidirectional Link Delay Sub-TLV.




Similarly as described in Section \ref{sub:igp}, we assigned Prefix-SID to FlexAlgo 130 on each router. Once the computation of FlexAlgo 130 was completed, we could see the paths to reach other nodes.




To steer the traffic within VRF BRONZE via paths computed by FlexAlgo 130 we associated FlexAlgo 130 with ODN color 30. We configured it similarly as we did with FlexAlgo 128.

As a last part of this experiment, we changed the delay on the link between router 2 and router 4 from 100 \textmu s to 10 \textmu s. We wanted to see if FlexAlgo would dynamically adapt its calculation to a changing environment. 

\subsection{Experiment 3 - TE metric}\label{sub:te}

In Experiment 3, we aimed to configure FlexAlgo 131 for VRF PLATINUM with computation determined by the TE Default metric. Since the underlying protocol responsible for calculating the Shortest Path Tree is OSPF, we wanted to show that load balancing would be possible on two equally costed paths. We began the configuration by specifying the default TE values on interfaces.


The routers advertised default TE metrics in the OSPFv2 Extended Link Opaque LSA of the OSPF LSU Packet. 
%
%
%
%
We assigned a Prefix-SID to FlexAlgo 131 on each router. 
As we aimed to demonstrate there are two possible paths to reach node 4.4.4.4 with an equal distance of 3. 
%
%
%
%
Lastly, we associated FlexAlgo 131 with ODN color 40 similarly as we did with FlexAlgo 128.

\subsection{Path Controller}\label{sub:pathcontroller}
To automate the injection of the routers' configurations related to FlexAlgo, we created an automation tool that could act as a part of the UPIN Path Controller \cite{UPIN3}. In particular, we aimed to automate the creation of FlexAlgo associated with VRF CUSTOM. In order to do so, we used the Cisco YANG suite to generate RPC payloads and Ansible to automate the whole procedure.

The automation process consisted of three steps. The first one was related to gathering input from the user with their desired FAD parameters. The user could specify three arguments, namely, one of the three administrative groups (include-any, exclude-any, include-all), an attribute color name related to this administrative group (red, blue), and a metric type (te-metric, delay, igp). However, the user did not have control over choosing FlexAlgo value as this could affect overriding the already existing FADs. That is why the tool automatically chose the ID of FlexAlgo.

We created an Ansible task that executed this operation on the routers using the NETCONF protocol. We gathered information about all FlexAlgos and we checked if the FlexAlgo chosen by a user already existed. If so, then the tool associated Prefix-SID of this exact FlexAlgo with the On-Demand policy for color 50 (VRF CUSTOM); if not, the tool creates this new FlexAlgo.

The configuration of the routers, the output of the \textit{show} commands, and the code for our configuration tool can be found in the following Github repository \cite{github}.

\section{Results}\label{Results}

For testing purposes, we attached the Ubuntu host to router 1 consecutively to each interface associated with different VRFs. We conducted two types of tests. As a first test, we run a traceroute to an IP address on router 4 from the same VRF. In the second test, we generated multiple TCP packets using a network tool called hping3. We performed both tests in three defined experiments. 

\subsection{Experiment 1 - results}\label{exp1-results}

First, we connected the Ubuntu host to router 1 interface GigabitEthernet0/0/0/5 which was associated with VRF GOLD. We added a 20.10.1.2 IP address on the interface facing router 1. 
We generated multiple TCP packets from the Ubuntu host. 
We sent TCP packets on port 80 to destination IP addresses from the 20.10.4.0/24 subnet. 
%
This command allowed us to send TCP packets on port 80 to IP addresses from the range 20.10.4.1-20.10.4.254. We wanted to generate multiple different flows to see if all traffic was forwarded via the same path. We captured the packets on all links using the feature provided by CML.
%
%
We also performed a traceroute command from router 1. 
%
%

\begin{figure}[t!]
    \centering
    \resizebox{\columnwidth}{!}{
    \includegraphics{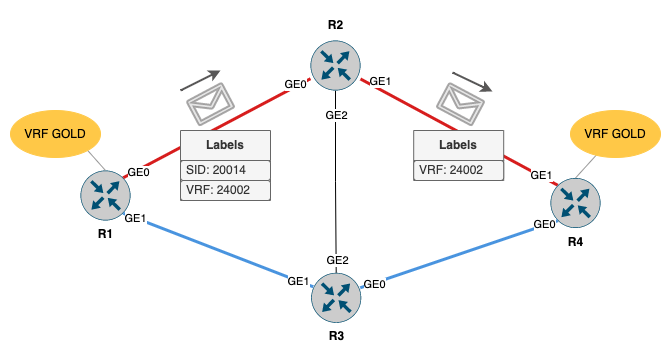}}
    \caption{Path chosen for FlexAlgo 128}
    \label{fig:algo128}
\end{figure}

From outputs of both traceroute and captured packets we conclude that the traffic for users from VRF GOLD was forwarded as shown in Figure \ref{fig:algo128}. We looked into the packets that were captured on the interfaces and we observed MPLS labels which are also presented in the figure. 

Router 1 created the ordered lists of instructions which are represented by the MPLS label stack. The label stack consisted of two MPLS labels, namely 20014, and 24002. The top one was associated with the Prefix-SID advertised by router 4 and associated with FlexAlgo 128. Router 2 popped this label from the stack when packets were forwarded to the next hop which was router 4. The bottom label 24002 was associated with VRF GOLD. 

We then connected Ubuntu to interface GigabitEthernet0/0/0/6 of router 1 which was associated with VRF SILVER. We addressed the Ubuntu host's interface facing router 1 with 20.20.1.2 IP. Again, we generated multiple TCP packets from the Ubuntu host, and in the same manner, we captured packets on all links to verify the selected path. 
We successfully tested the functionality of FlexAlgo 129 in a similar way, this time avoiding red links.

\subsection{Experiment 2 - results}\label{exp2-results}

We attached the Ubuntu host to router 1 interface GigabitEthernet0/0/0/7 which was associated with VRF BRONZE. We addressed the interface on Ubuntu that was facing router 1 with the 20.30.1.2 IP. Then, we performed a traceroute to the 20.20.4.4 IP address. We generated TCP packets to multiple destination IP addresses from subnet 20.30.4.0/24 and we captured the traffic on all links. In the meantime, we changed the delay on the link between router 2 and router 4 from 100 \textmu s to 10 \textmu s.
The output of the traceroute demonstrated how many packets were sent on each link from the source and destination located in VRF BRONZE.

After, we changed the delay on the link between router 2 and router 4 from 100 \textmu s to 10 \textmu s we also performed the traceroute again.





As we can observe from the results of both tests, routers dynamically adjusted the computation of FlexAlgo after the delay changed. 

\subsection{Experiment 3 - results}\label{exp3-results}

We connected the Ubuntu host to the router 1 interface GigabitEthernet0/0/0/8, which was associated with the VRF PLATINUM. We assigned the 20.30.1.2 IP address to the interface on the Ubuntu host, facing router 1. Then, we ran a traceroute to the 20.40.4.4 IP address. We wanted to generate multiple packet flows to see if the traffic was load-balanced on equally costed paths.
We sent packets from multiple spoofed IP addresses from subnet 20.40.1.0/24 to multiple destination IP addresses from the 20.30.4.0/24 subnet. 





\begin{figure}[t!]
    \centering
    \resizebox{\columnwidth}{!}{
    \includegraphics{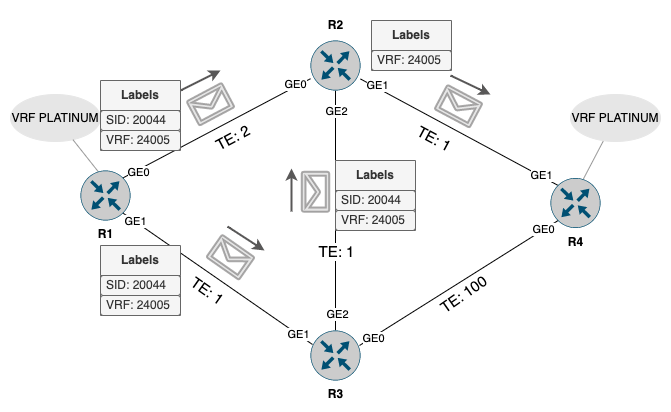}}
    \caption{Path chosen for FlexAlgo 131}
    \label{fig:algo131}
\end{figure}

The path chosen by the routers for FlexAlgo 131 is illustrated in Figure \ref{fig:algo131}. Additionally, the figure shows the MPLS labels from the generated packets. Label 20044 corresponded to the Prefix-SID associated with FlexAlgo 131 advertised by router 4, while label 24005 corresponded to the VRF PLATINUM destinations attached to router 4.


The results of running traceroute and capturing the generated packets demonstrate that traffic was load balanced between two equally costed paths computed by FlexAlgo 131.

\subsection{Path Controller - results}
As stated in Section \ref{sub:pathcontroller}, in order to test the Path Controller, we performed two tests that imitated two different scenarios.

In the first scenario, the user specified the FlexAlgo with the same parameters as an already existing FlexAlgo 128, namely \textit{igp} as metric type,  \textit{exclude-any} as affinity, and \textit{blue} as the attribute related to this affinity. The user inserted these values through a command-line user interface. One of the tasks within our tool compared the values specified by the user with other FlexAlgos parameters stored in a dictionary. It found a match with a FlexAlgo with ID 128.
Therefore, it did not create a new FlexAlgo but associated the ID of FlexAlgo 128 with the ODN color 50. This resulted in injecting a new configuration to router 1 and router 4.

In the second scenario, the user specified the FlexAlgo with parameters that did not match with any other FlexAlgo stored in a dictionary, namely \textit{te-metric} as metric type,  \textit{include-all} as affinity, and \textit{red} as the attribute related to this affinity. The following FAD did not match with any of the FlexAlgos, therefore the automation tool created a new FAD, specified a new Prefix-SID for this FlexAlgo on all routers and finally associated the ID of a new FlexAlgo with the VRF CUSTOM on router 1 and router 4. 
As the highest value of the already existing FlexAlgos was 131, the tool automatically assigned the FlexAlgo ID of value 132 to the newly created FlexAlgo. 

\section{Conclusions} \label{Discussion and Conclusions}

In this work we set out to test the suitability of FlexAlgo as one of the technologies that can underpin the Responsible Internet. We proved it to be a scalable technique as it requires only a single Prefix-SID to enforce traffic to specific paths, without the need to advertise Prefix-SIDs for the transit nodes.
Moreover, FlexAlgo is flexible as it enables the integration of multiple unique algorithms within a single SR domain. It enables the creation of multiple logical topologies by assigning a single FlexAlgo ID to one set of routers. 
Another advantage is that FlexAlgo benefits from the simplicity and automation features of the SR-TE solutions. 
Finally, the configuration of FlexAlgo can be automated with NETCONF and Yang modules. 


When FlexAlgo uses the SR-MPLS control plane, as we tested in our environment, Prefix-SIDs characterized as MPLS labels are distributed within the IGP Link State packets. This way, FlexAlgo does not require any additional label distribution or path signaling protocol, and thus, generates less overhead compared to traditional MPLS solutions.

Furthermore, FlexAlgo can help fulfilling one of the main goals of the UPIN project by handing over some control over the traffic to the user to suit their specific needs.
With FlexAlgo, users could experience greater visibility and control over their traffic, leading to a more trustworthy network environment. Specifically, with the automation tool, we demonstrated a way for users to set up a path following constraints they chose. They are able to choose more secured paths based on their specific needs through constraints they set themselves. 
This allowed us to open an interface between the network, its routing functionalities, and its users, therefore not offering routing as a black box but instead provide users with insight and more control compared to traditional approaches. This increases the situational awareness and provides an additional layer of security and privacy for user's data in transit.



The limitations we came across when performing our experiments were related to our virtualized environment, CML, where IOS XR routers did not support real-time delay measurement. As a workaround, we had to statically advertise delay on links.
Another limitation that we came across was related to FlexAlgo itself and TE metrics. 
The draft \cite{Psenak_Hegde_Filsfils_Talaulikar_Gulko_2022} does not specify how exactly TE metric extensions could be introduced and implemented within FlexAlgo. Furthermore, we did not find how TE metric extensions could be employed with FlexAlgo on IOS XR routers.

We were able to test the capabilities and limitations of FlexAlgo in a confined environment.
We plan to carry out experiments with FlexAlgo on a more complex and physical network topology, as well as on IS-IS-based control plane and SRv6 based data plane to investigate the differences in both configuration and capabilities of FlexAlgo.

\section*{Acknowledgements}

This research received funding from the Dutch Research Council (NWO) under the project UPIN and the project CATRIN under the Dutch Research Agenda (NWA) – Cybersecurity - Research Programme.

\printbibliography
\end{document}